\documentstyle[12pt,epsf]{article}

\begin{document}
 
\begin{titlepage}

\title{
\hfill
\parbox{5cm}{\normalsize 
% ICRR-Report-???-??-??\\
\normalsize UT-808\\
}\\
\vspace{2ex}\large
Investigation into O(N) Invariant Scalar Model Using Auxiliary-Mass Method 
at Finite Temperature
\vspace{2ex}}
\author{\large
Kenzo Ogure\thanks{e-mail address:
 {\tt ogure@icrhp3.icrr.u-tokyo.ac.jp}}\\
 {\it Institute for Cosmic Ray Research,
   University of Tokyo, Midori-cho,}\\
   {\it   Tanashi, Tokyo 188, Japan}\\
and\\
Joe Sato\thanks{e-mail address:
 {\tt joe@hep-th.phys.s.u-tokyo.ac.jp}}\\
 {\it Department of Physics, School of Science,
   University of Tokyo,}\\
   {\it   Tokyo 113, Japan}}
\date{\today}

\maketitle

\begin{abstract}
  Using auxiliary-mass method, O(N) invariant scalar model is
  investigated at finite temperature. This mass and an evolution
  equation allow us to calculate an effective potential without an
  infrared divergence. Second order phase transition is indicated by
  the effective potential. The critical exponents are determined
  numerically.
\end{abstract}

\end{titlepage}

\newpage

Symmetry restoration of O(N) scalar model at high temperature is very
important since many physical systems belong to the same universality
class as it: polymer phase transition (N$\rightarrow$0), critical
liquid-vapor phase transition (N=1), alloy (e.g. $\beta$-brass) phase
transition (N=1), uniaxial ferromagnet phase transition (N=1),
superfluid phase transition (N=2), ferromagnet phase transition (N=3),
and chiral phase transition with two flavor massless quarks (N=4)
\cite{ZJ,Wil}.\\

The phase transition had better be investigated by finite-temperature
field theory, which is based on the statistical principle only.  The
perturbation theory, however, breaks down around the critical
temperature when the phase transition is second or weakly first order
\cite{Arn2}, even if daisy-diagrams (ring-diagrams) \cite{Kap,Tak,Fen}
are resumed. Investigation into phase transitions at
finite-temperature has long been hampered by this failure. Many
methods to avoid the failure were proposed: CJT method \cite{Ame},
renormalization improvement \cite{Nak}, novel summation \cite{Chi},
pad\'{e} improvement \cite{Hat}, exact renormalization group with
temperature \cite{Lia}, and auxiliary-mass method \cite{DHL,A,C}.  \\ 

The auxiliary-mass method is used in the present paper. The method is
based on the following idea.  First an effective potential can be 
calculated at large mass by the perturbation theory since it is
reliable there.  Second the effective potential is extrapolated to the
small mass range, where the perturbation theory is not reliable, using
an evolution equation. Finally various quantities (e.g. critical
exponents) are determined from the effective potential.  The method is
applied to O(N) scalar model concretely in the following; the phase
transition of the model is investigated and the critical exponents are
determined.\\ 

We consider the following Lagrangian density,
\begin{eqnarray}
     {\cal L}_{E}=-\frac{1}{2}
     \left(\frac{\partial \phi_a}{\partial \tau}\right)^{2}
     -\frac{1}{2}(\mbox{\boldmath $\nabla$} \phi_a)^{2}
     -\frac{1}{2}m^{2}\phi_a^{2}
     -\frac{\lambda}{4!}(\phi_a^{2})^{2}
     +J_a\phi_a + c.t.\ \ ,
    \label{lag}
\end{eqnarray}
where $J_a$ is an external source function and the index ``$a$'' runs
from 1 to N.  First the effective potential $V$ is calculated by the
perturbation theory within one-loop order at large mass $m^2=
M^2=O(T^2)$. We choose field expectation values $\bar\phi_a=\bar\phi \ 
 \delta_{1a}$ without a loss of generality because of O(N) invariance:
\begin{eqnarray}
     V&=&\frac{1}{2}M^{2}\bar{\phi}^{2}
     +\frac{\lambda}{4!}\bar{\phi}^{4}
     +\frac{T}{2\pi^{2}}
     \int^{\infty}_{0}dr r^{2}\log \left[
     1-\exp\left(-\frac{1}{T}\sqrt{r^{2}+M^{2}+
     \frac{\lambda}{2}\bar\phi^{2}}\right)\right]\nonumber \\
     &&+(N-1)\ \frac{T}{2\pi^{2}}
     \int^{\infty}_{0}dr r^{2}\log \left[
     1-\exp\left(-\frac{1}{T}\sqrt{r^{2}+M^{2}+
     \frac{\lambda}{6}\bar\phi^{2}}\right)\right].\label{pot}
\end{eqnarray}
We note that the daisy-resummation is not necessary because of the
large mass and one-loop zero-temperature effect is negligible if the
coupling is weak. Next this effective potential is extrapolated to
smaller mass using the following evolution equation \cite{A},
\begin{eqnarray}
     \frac{\partial V}{\partial m^{2}}&=&
     \frac{1}{2}\bar{\phi}^{2}+\frac{1}{4\pi i}
     \int^{+i\infty +\epsilon}_{-i\infty +\epsilon}dp_{0}
     \int\frac{d^{3}\mbox{\boldmath $p$}}{(2\pi)^{3}}
     \frac{1}{-p_{0}^{2}+\mbox{\boldmath $p$}^{2}+m^{2}
     +\frac{\lambda}{2}
     \bar{\phi}^{2}+\Pi_{Sig}}\frac{1}{e^{\frac{p_{0}}{T}}-1}\nonumber \\
     &&+\frac{N-1}{4\pi i}
     \int^{+i\infty +\epsilon}_{-i\infty +\epsilon}dp_{0}
     \int\frac{d^{3}\mbox{\boldmath $p$}}{(2\pi)^{3}}
     \frac{1}{-p_{0}^{2}+\mbox{\boldmath $p$}^{2}+m^{2}
     +\frac{\lambda}{6}
     \bar{\phi}^{2}+\Pi_{NG}}\frac{1}{e^{\frac{p_{0}}{T}}-1}.\nonumber \\
    \label{evo}
\end{eqnarray}
Here $\Pi_{Sig}=\Pi_{Sig}(m^2,\bar\phi^2,p_0^2,{\bf p^2},T)$ and
$\Pi_{NG}=\Pi_{NG}(m^2,\bar\phi^2,p_0^2,{\bf p^2},T)$ are the full
self-energy, which are those for massive and massless mode in the broken
phase.  This equation is modified from that of $\lambda\phi^4$ theory
\cite{A} straitforwardly through a diagonalization of the propagator.
Though this equation is exact, it can not be solved without an
approximation; because it includes the full propagator which can not
be known exactly.  We then replace as follows,
\begin{eqnarray}
     m^{2}+\frac{\lambda}{2}\bar{\phi}^{2}+
     \Pi_{Sig}(0,0,\bar{\phi},m^{2},\tau)\rightarrow
     \frac{\partial^{2}V}{\partial\bar\phi^{2}}\nonumber\\
     m^{2}+\frac{\lambda}{6}\bar{\phi}^{2}+
     \Pi_{NG}(0,0,\bar{\phi},m^{2},\tau)\rightarrow
     \frac{1}{\bar\phi}
     \frac{\partial V}{\partial\bar\phi}.
    \label{rep}
\end{eqnarray}
This replacement corresponds to fixing an external momentum of the
full self-energy to zero\footnote{
This is the first approximation of a systematic calculation \cite{E}.
}.  
It allows us to convert eq.(\ref{evo}) to
the following partial differential equation,
\begin{eqnarray}
     \frac{\partial V}{\partial m^{2}}&=&
     \frac{1}{2}\bar{\phi}^{2}+\frac{1}{4\pi^{2}}
     \int^{\infty}_{0}dr r^{2}\frac{1}{\displaystyle \sqrt{r^{2}
     +\frac{\partial^{2}V}{\partial\bar\phi^{2}}}}
     \frac{1}{\displaystyle \exp\left(\frac{1}{T}\sqrt{r^{2}
     +\frac{\partial^{2}V}{\partial\bar\phi^{2}}}\right)-1}\nonumber\\
     &&+\frac{N-1}{4\pi^{2}}
     \int^{\infty}_{0}dr r^{2}\frac{1}{\displaystyle \sqrt{r^{2}
     +\frac{1}{\bar\phi}\frac{\partial V}{\partial\bar\phi}}}
     \frac{1}{\displaystyle \exp\left(\frac{1}{T}\sqrt{r^{2}
     +\frac{1}{\bar\phi}
     \frac{\partial V}{\partial\bar\phi}}\right)-1}.
    \label{evo2}
\end{eqnarray}
We numerically solve the equation to $m^2=-\mu^2$, where O(N) symmetry
is broken at zero-temperature, under the initial condition
eq.(\ref{pot}). We solve it above the critical temperature only and can
get sufficient information about the critical phenomenon.\\

\begin{figure}
\unitlength=1cm
\begin{picture}(16,6)
%\input zahyou
%\zahyou{16}{6}
\unitlength=1mm
\put(15,35){$\displaystyle  V$}
\put(120,0){$\displaystyle \bar\phi$}
\centerline{
\epsfxsize=10cm
\epsfbox{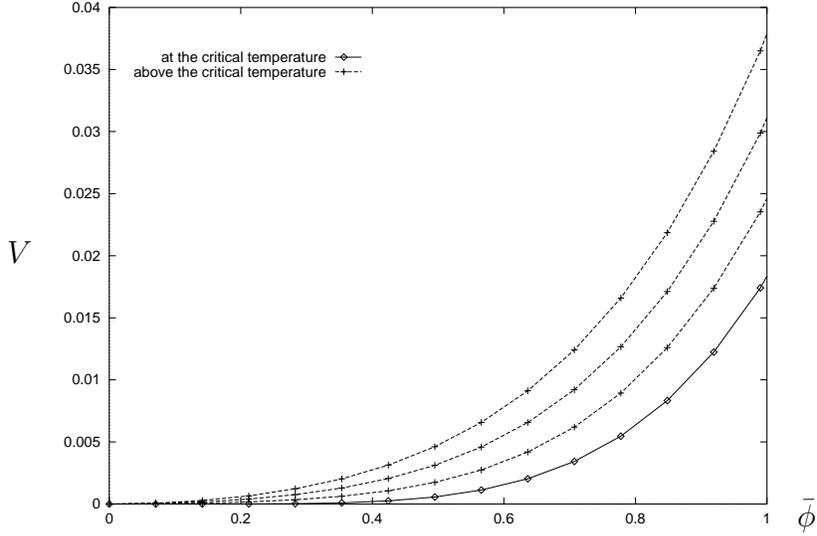} 
} 
\end{picture}
\caption{The effective potential obtained by the auxiliary-mass
  method ($N=4,\lambda =1$). Second-order phase transition occurs at
  the critical temperature. Similar behavior are observed for the
  other N and $\lambda$.}
\label{cripot}
\end{figure}

The effective potential are calculated using numerical methods in
Ref.\cite{C}.  They are shown in fig.\ref{cripot}.  We can not observe
a second minimum which is a hallmark of a first order phase
transition. Instead, the curvature of the origin decreases and
vanishes smoothly as temperature decreases; one can, therefore,
observe that the phase transition of the model is second order as it
should be.\\ 

\begin{figure}
\unitlength=1cm
\begin{picture}(16,6)
%\input zahyou
%\zahyou{16}{6}
\unitlength=1mm
\put(15,35){$\displaystyle  T_c/\mu$}
\put(120,0){$\displaystyle N$}
\centerline{
\epsfxsize=10cm
\epsfbox{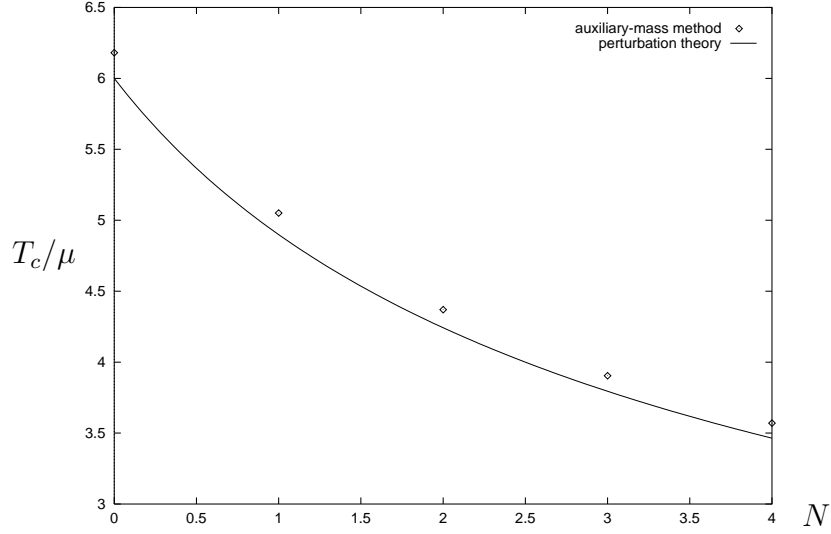} 
} 
\end{picture}
\caption{Critical temperature as a function of N at $\lambda=1$
($\diamond$). This
resembles to a result of the perturbation theory at leading order
(-----) 
%, $T_c=6 \sqrt\frac{2}{\lambda (N+2)}$, 
but has a slight difference quantitatively.}
\label{critem}
\end{figure}

The critical temperature is shown as a function of N in
fig.\ref{critem}. Though they resemble to a result of the perturbation
theory at leading order \cite{Dol}, $T_c=6 \sqrt\frac{2}{\lambda (N+2)}$, which is
determined from the condition that mass with the daisy-diagram
vanishes, they have a slight difference quantitatively.\\ 

\begin{figure}
\unitlength=1cm
\begin{picture}(16,6)
%\input zahyou
%\zahyou{16}{6}
\unitlength=1mm
\put(-10,35){$\displaystyle 
  \log[(\frac{\partial^2 V}{\partial\bar\phi^2})/\mu^2]$}
\put(120,0){$\displaystyle \log\tau$}
\centerline{
\epsfxsize=10cm
\epsfbox{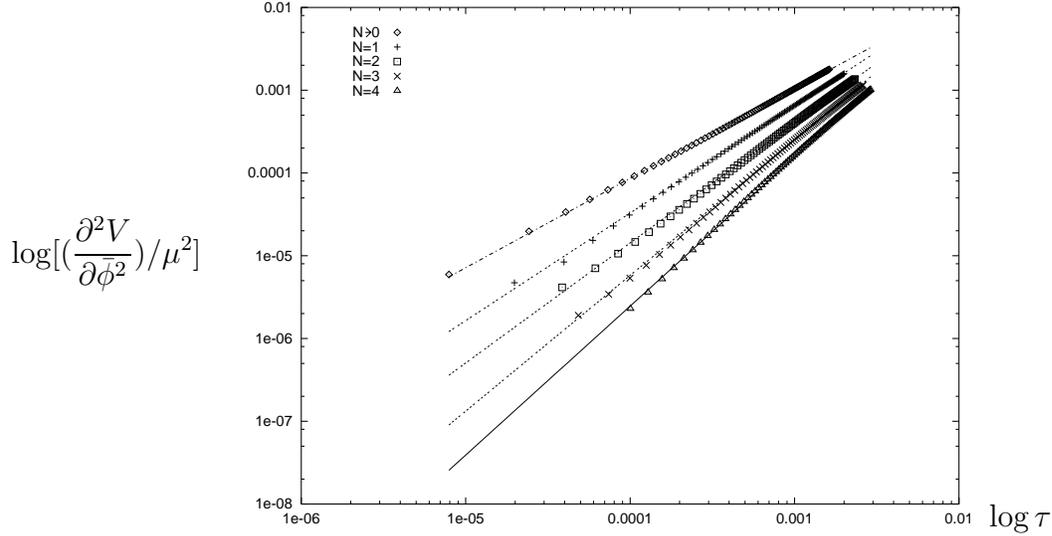} 
} 
\end{picture}
\caption{Second derivative of the effective potential with respect to
$\phi$.  The gradients are steeper for larger N. }
\label{cur}
\end{figure}

Finally the critical exponents, $\gamma$ and $\delta$, are determined
using this potential.  We determine $\gamma$ from a second derivative
of the effective potential with respect to $\bar\phi$ at the origin. 
The critical exponent $\gamma$ is defined as follows,
\begin{eqnarray}
    \chi
    &\equiv&
    \left.
    \frac{\partial\bar\phi}{\partial J_1}
    \right|_{J_a=0}
    \sim\chi\tau^{-\gamma}  \ \ \   (\tau =(T-T_{c})/T_c)
    \label{gamma}
\end{eqnarray}
the following identity then relates the second derivative to the
susceptibility,  
\begin{eqnarray}
    \label{id}
    \left.
    \frac{\partial\bar\phi}{\partial J_1}
    \right|_{J_a=0}
    =
    \left.
    \left(
    \frac{\partial^2 V}{\partial \bar\phi^2}
    \right)^{-1}
    \right|_{\bar\phi=\phi_c}.    
\end{eqnarray}
Here $\phi_c$ is determined from the condition, $\left.\frac{\partial
V}{\partial \bar\phi}\right|_{\bar\phi=\phi_c}=0$. The second
derivative are shown in log-scale in fig.\ref{cur}. We then determine
the gradients, which is the very $\gamma$ we want. One can observe
that they become steeper as N increases; then, $\gamma$ become larger.
The results are $\gamma =\ 1.13 \ (N\rightarrow 0),\ 1.37 \ (N=1), \ 
1.47 \ (N=2),\ 1.60 \ (N=3),\ 1.66 \ (N=4)$. These results are
summarised in table.(\ref{com}), compared with the Landau approximation
and world best values.\\ 

We determine $\delta$ using the effective potential at the critical
temperature.  The critical exponent $\delta$ is defined as follows.  
\begin{eqnarray}
    \label{delta}
    \bar\phi\propto J_1^{1/\delta}=(\frac{\partial
    V}{\partial\bar\phi})^{1/\delta} \hspace{1cm}(T=T_c).
\end{eqnarray}
The following relation, which derived from eq.(\ref{delta}), enables us to
determine $\delta$ from the effective potential at the critical
temperature,
\begin{eqnarray}
    \label{id2}
    V\propto\bar\phi^{\delta+1}\hspace{1cm}(T=T_c).
\end{eqnarray}
The effective potential at the critical temperature is shown in
fig.\ref{cripot}.  We determine $\delta$ from it.  The result are
$\delta = \ 3.8\ (N \rightarrow 0),\ 4.0\ (N=1),\ 4.2\ (N=2),\ 
4.4\ (N=3),\ 4.4\ (N=4)$.\\

%The results are summarised in table.\ref{result}.
\begin{table}[ht]
{\small
\begin{center}
\begin{tabular}{|l|c|c|}
\hline
    & $\gamma$ (LA,WBV)   & $\delta$ (LA,WBV) \\
\hline
N$\rightarrow$0 \cite{ZJ}&    1.13 (1, 1.16)   &  3.8  (3, 4.77)  \\
\hline
N$=$1 \cite{ZJ}&    1.37  (1, 1.24)  &   4.0  (3, 4.76) \\
\hline
N$=$2 \cite{ZJ}&   1.47   (1, 1.32)  &   4.2   (3, 4.78)\\
\hline
N$=$3 \cite{ZJ}&   1.60    (1, 1.40) &   4.4  (3, 4.78) \\
\hline
N$=$4 \cite{Kan} &   1.66    (1, 1.48) &  4.4  (3, 4.85)  \\
\hline
\end{tabular}\\
\end{center}
}
\caption{The critical exponents, $\gamma$ and $\delta$, obtained in
the present paper. Those of Landau approximation (LA) and world best
values (WBV) are also summarised. We used lattice results as WBV. }
\label{com}
\end{table}

In the present paper, we showed that the phase transition of O(N)
scalar model is second order using auxiliary-mass method at finite
temperature.  This is a great progress because we can not show it
using the perturbation theory with daisy-resummation ------ the
traditional method to calculate the effective potential at finite
temperature------ \cite{Arn2}.\\ 

Since the phase transition turned out to be second order, we
determined the critical exponents of the phase transition. Though
the results are not as accurate as world best values, they are much
better than that obtained in the Landau approximation.  The error would be 
due to the replacement (\ref{rep}). We must improve the approximation 
in order to get more accurate values \cite{E}.\\

In conclusion, the auxiliary-mass method turns out to be reliable not
only qualitatively but also in rough quantitative estimation in O(N)
invariant scalar model. This method would be reliable in the other
models. What is more, we can investigate not only second-order phase
transition but also first-order phase transition without any
modifications \cite{B}. This method, therefore, enable us to investigate the
phase transition of various models: the cubic anisotropy model, the abelian
Higgs model, and the standard model. \\

We finally express our thanks to T. Inagaki for valuable discussions
and communications. J.S. is supported by JSPS fellowship.

\end{document}